\documentclass[acmsmall]{acmart}

\AtBeginDocument{%
  \providecommand\BibTeX{{%
    \normalfont B\kern-0.5em{\scshape i\kern-0.25em b}\kern-0.8em\TeX}}}

\usepackage{caption}
\usepackage{subcaption}
\usepackage{xcolor}
\usepackage{tabularx}
\usepackage{multirow}

\usepackage{pgfplots}
\usepackage{arydshln}
\usepackage{comment}

\begin{document}

\title{Uncovering CWE-CVE-CPE Relations with  Threat Knowledge Graphs}

\author{Zhenpeng Shi}
\affiliation{%
  \institution{Boston University}
  \city{Boston}
  \state{MA}
  \country{USA}}
\email{zpshi@bu.edu}

\author{Nikolay Matyunin}
\affiliation{%
  \institution{Honda Research Institute Europe GmbH}
  \city{Offenbach am Main}
  \country{Germany}}
\email{nikolay.matyunin@honda-ri.de}

\author{Kalman Graffi}
\affiliation{%
  \institution{Technische Hochschule Bingen}
  \city{Bingen}
  \country{Germany}}
\email{k.graffi@th-bingen.de}

\author{David Starobinski}
\affiliation{%
  \institution{Boston University}
  \city{Boston}
  \state{MA}
  \country{USA}}
\email{staro@bu.edu}


\begin{abstract}
Security assessment relies on public information about products, vulnerabilities, and weaknesses. 
So far, databases in these categories have rarely been analyzed in combination. 
Yet, doing so could help predict unreported vulnerabilities and identify common threat patterns.
In this paper, we propose a methodology for producing and optimizing a knowledge graph that aggregates knowledge from common threat databases (CVE, CWE, and CPE).
We apply the threat knowledge graph to predict associations between threat databases, specifically between products, vulnerabilities, and weaknesses.
We evaluate the prediction performance both in closed world with associations from the knowledge graph, and in open world with associations revealed afterward. 
Using rank-based metrics (i.e., Mean Rank, Mean Reciprocal Rank, and Hits@N scores),
we demonstrate the ability of the threat knowledge graph to uncover many associations that are currently unknown but will be revealed in the future, which remains useful over different time periods.
We propose approaches to optimize the knowledge graph, and show that they indeed help in further uncovering associations.
\end{abstract}



\maketitle

\section{Introduction}

Security assessment relies on public knowledge about existing vulnerabilities. For instance, Software Composition Analysis (SCA) scanners, such as the OSV-Scanner~\cite{osv}, rely on knowledge about disclosed vulnerabilities to identify vulnerable dependencies in the codebase. Likewise, static Application Security Testing (SAST) tools  utilize categorization knowledge to provide unified reports about discovered issues, and Dynamic Application Security Testing (DAST) scanners dynamically probe applications for known vulnerabilities. 
Several threat modeling tools (e.g., OWASP pytm, Threagile, IriusRisk) use publicly disclosed weaknesses as pre-defined threats in their libraries~\cite{shi21taxonomy}. By checking whether the conditions of said pre-defined threats are met, these tools can identify potential threats in a system, for which corresponding mitigations can be developed.

The security assessment activities mentioned above rely on \emph{threat databases} that provide useful common knowledge on vulnerabilities, weaknesses, and products.
Prominent examples include the Common Vulnerabilities and Exposures (CVE)~\cite{cve},  Common Weakness Enumeration (CWE)~\cite{cwe}, and Common Platform Enumeration (CPE)~\cite{cpe} databases.
The CVE lists publicly disclosed security vulnerabilities. Presently, it aggregates over 200,000 entries, with each CVE entry containing a short description and references to the disclosure reports. The CWE lists more than 900 generic software and hardware weakness types. CWE entries classify general instances of a vulnerability. For example, CWE-502 corresponds to a software weakness that deserializes untrusted data without sufficient verification. This entry relates to many examples being discovered and disclosed as CVE entries for specific products (e.g., CVE-2021-21348, which affects the XStream library).
The CWE further provides a categorization of weaknesses into \emph{views} and \emph{categories}~\cite{view}: a category contains a set of weaknesses that share a common characteristic, and a view provides a specific perspective of examining CWE contents, such as weaknesses in software written in C.
Last, the CPE provides a structured naming scheme for common products, including information technology systems, software, and packages, and lists the names of common products following the scheme.
\definecolor{bblue}{RGB}{0,114,209}

\begin{table}[t]
    \centering
    \begin{tabular}{|p{0.25\linewidth}|p{0.64\linewidth}|}
    \hline
        CVE ID & CVE-2021-21348 \\ \hline
        Associated CWE & CWE-400, CWE-502 \\
        \hline
        Description & XStream is a Java library to serialize objects to XML and back again. In XStream before version 1.4.16, there is a vulnerability which may allow a remote attacker to occupy a thread that consumes maximum CPU time and will never return. ... \\
        \hline
        Associated CPE \newline by Aug 4, 2021  &   1) cpe:a:xstream\_project:xstream:*:*, \newline
        2) cpe:o:debian:debian\_linux:*:* \\ \hline
        Associated CPE \newline after Aug 4, 2021 \newline &  1) \textcolor{bblue}{cpe:o:fedoraproject:fedora:*:*},\newline
        2) \textcolor{bblue}{cpe:a:oracle:retail\_xstore\_point\_of\_service:*:*},\newline
        3) \textcolor{bblue}{cpe:a:oracle:webcenter\_portal:*:*},\newline
        4) \textcolor{bblue}{cpe:a:oracle:banking\_platform:*:*},\newline
        5) \textcolor{bblue}{cpe:a:oracle:communications\_policy\_management:*:*}\newline
        6) \textcolor{bblue}{cpe:a:oracle:communications\_billing\_and\_\newline revenue\_management\_elastic\_charging\_engine:*:*}, \newline
        7) \textcolor{bblue}{cpe:a:oracle:mysql\_server:*:*},\newline
        8) \textcolor{bblue}{cpe:a:oracle:business\_activity\_monitoring:*:*},\newline
        9) cpe:a:oracle:communications\_unified\_inventory\_\newline  management:*:*,\newline
        10) cpe:a:oracle:banking\_virtual\_account\_management:*:*
         \\ \hline
    \end{tabular}
    \caption{Example of prediction results for CVE-2021-21348. The products are listed in the form of {\tt cpe:part:vendor:product:target software:target hardware}. 
    Using a threat knowledge graph generated with data available on August~4, 2021, we predict associations between CPE entries and the aforementioned vulnerability.
    8 out of 10 products are successfully predicted (marked in blue),
    with one false positive prediction.
    \vspace{-0.5cm}}
    \label{CVE-2021-21348}
\end{table}

The National Vulnerability Database (NVD)~\cite{nvd} provides associations between entries of the aforementioned databases. Table~\ref{CVE-2021-21348} shows an example of associations between the entries of the different threat databases~\cite{nvd-cve-2021-21348}. Specifically, vulnerability CVE-2021-21348 is an instance of weaknesses CWE-400 and CWE-502. Its exploitation affects various products, as specified by associated CPE entries.
Given software configurations or packages used in a system,  associations by NVD can suggest relevant potential vulnerabilities and weaknesses. These associations are  highly valuable to vulnerability scanners.
Furthermore, as threat modeling often involves different abstraction levels of a system,  CVE-CWE associations assist in mapping concrete threats (e.g., vulnerabilities in specific products) to abstract threats (e.g., general weaknesses) that can be used in higher-level threat models.

The creation of associations by the NVD is a lengthy process, since these associations are manually analyzed and vetted by security experts from credible organizations or corporations.
As such, at any given time, associations provided by the NVD may be incomplete. As an illustration, in Table~\ref{CVE-2021-21348}, we split the affected products of CVE-2021-21348 into two groups: those revealed before and after August~4, 2021. By August~4, 2021, only two CPE entries were known to be affected according to the NVD. However, many more affected entries were subsequently discovered and added. 
Being able to predict some of these associations in advance would be highly valuable, for instance, to identify and mitigate hidden vulnerabilities and weaknesses in a timely fashion.

\begin{figure*}[t]
	\centering
	\includegraphics[width=\linewidth]{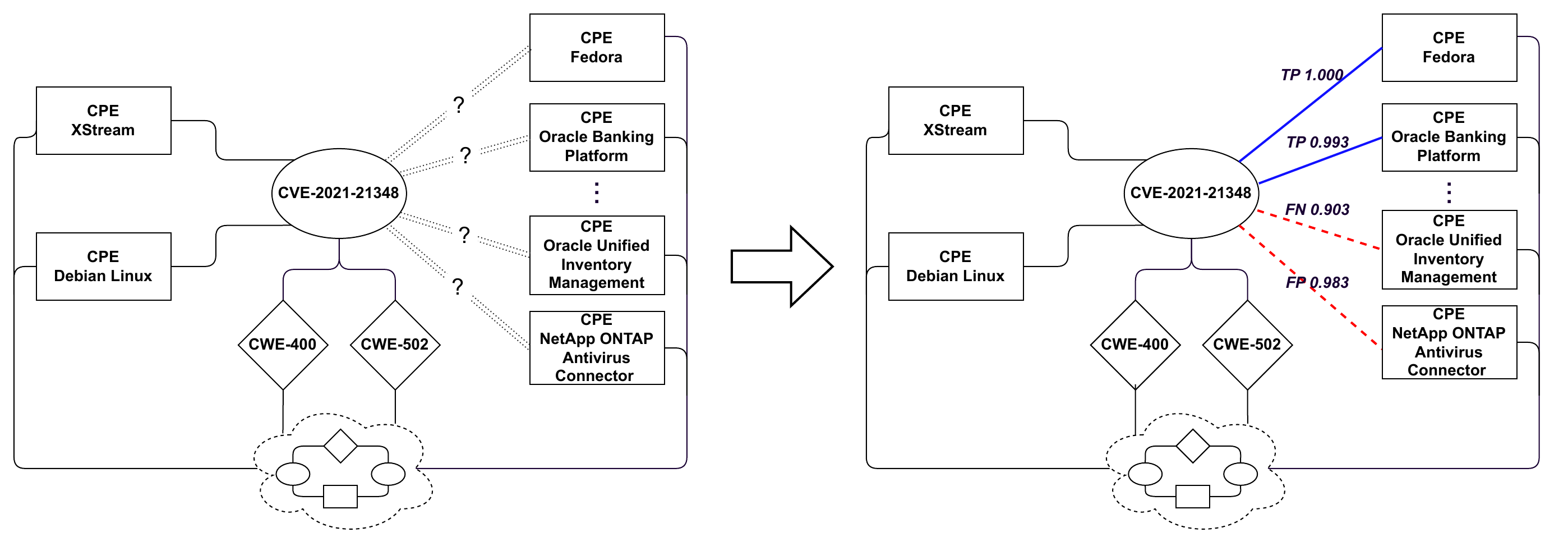}
	\caption{Illustrations of the prediction process. For any given CVE (e.g., CVE-2021-21348), and its known associations to CPEs  and CWEs at a certain date (left side), we aim to predict future associations to other CPEs and CWEs (right side). The cloud at the bottom represents the rest of CPE/CVE/CWE entries and the associations between them. The blue lines represent successfully predicted associations (true positives), and the red dashed lines represent failed predictions due to false positives or false negatives.
    }\label{example-kg}
\end{figure*}

In this paper, we propose a concrete methodology to facilitate such predictions. Our approach is based on the analysis of associations between entries of the various threat databases. To enable such analysis, we produce a knowledge graph~\cite{paulheim2017knowledge,chen2020review,ji2021survey} out of existing databases (CVE, CWE, and CPE), which we refer to as a \emph{threat knowledge graph}. We use the threat knowledge graph to predict future associations between threat database entries, as illustrated in Fig.~\ref{example-kg}.

We achieve association prediction by applying a machine learning method known as \emph{knowledge graph embedding}. As a result of the embedding process, a trained model can assign a score to each given association that indicates how likely it is for this association to belong to the knowledge graph. By enumerating over non-existent associations and ranking the resulting assigned scores, we can predict top candidates for future associations.
%
%
Table~\ref{table_ranking} illustrates this process.  Five triples are ranked by the probabilities assigned by the trained model and separated by a threshold.
Accordingly, the model predicts that associations above the threshold will be revealed in the future. In this paper, we conduct extensive evaluations using well-established metrics for knowledge graphs and show that this model indeed performs well in predicting threat associations.

\begin{table}[t]
\begin{tabular}{ccccc}
\hline
CPE-ID                             & CVE-ID         & Score    & Probability & Positivity \\ \hline
cpe:a:oracle:banking\_platform:*:* & CVE-2020-35491 & 8.656979 & 0.999826    & 1          \\ 
cpe:a:oracle:webcenter\_portal:*:* & CVE-2021-21348 & 5.275653 & 0.994911    & 1          \\ 
cpe:a:oracle:banking\_platform:*:* & CVE-2021-21348 & 4.921300 & 0.992763    & 1          \\ \hdashline
cpe:a:oracle:banking\_platform:*:* & CVE-2020-3990  & 1.521459 & 0.820753    & 0          \\ 
cpe:a:oracle:webcenter\_portal:*:* & CVE-2021-31874 & 0.303383 & 0.575270    & 0          \\ \hline
\end{tabular}
\caption{Illustration of threat knowledge graph ranking positive (existent) and negative (non-existent) associations. The predicted probabilities of the associations are normalized from their scores. 
By setting a proper threshold (represented by the dashed line), the five example associations can be successfully separated. 
}
\label{table_ranking}
\end{table}

Our work therefore makes the following contributions:
\begin{enumerate}
    \item We propose and implement the concept of \emph{threat knowledge graph} to aggregate knowledge from common threat databases. To do so:
    \begin{itemize}
        \item We translate  entries in threat databases and their associations into triples that form the knowledge graph. 
        \item We optimize the knowledge graph to improve its prediction capabilities.
    \end{itemize}

    \item We embed the knowledge graph onto a vector space that can be used for link prediction. More specifically:
    \begin{itemize}
        \item We compare several embedding models (i.e., TransE, DistMult, and ComplEx). We show that the TransE model consistently performs the best for the task at hand.
        \item We perform  a \emph{closed-world} evaluation of the embedded knowledge graph using standard rank-based metrics, including the Mean Rank, the Mean Reciprocal Rank, and Hits@N scores. 
        The evaluation demonstrates the high quality of the embedding (e.g., achieving a 0.606 Hits@10 score in predicting CVE-CPE associations).
    \end{itemize}

    \item We further evaluate  the ability of threat knowledge graphs to predict in an \emph{open-world} setting associations between the entries in threat databases (i.e., between products, vulnerabilities, and weaknesses):
    \begin{itemize}
        \item We use the embedded knowledge graph to predict new associations that were discovered after August~4, 2021.
        \item We evaluate the prediction results using rank-based metrics as well as precision, recall, and F1-score. The metrics suggest that the model can make useful predictions (e.g., a 0.358 Hits@10 score in predicting newly added CVE-CWE associations, and an F1-score of 0.681 for CVE-CPE predictions).
        \item We demonstrate the usefulness of the prediction capability over different time periods, by producing and evaluating a threat knowledge graph from data available on March~29, 2022.
    \end{itemize}

    \item We investigate approaches to further improve the quality of the knowledge graph, specifically by removing obsolete entries and incorporating data from other databases (i.e., CAPEC entries and CVSS vectors). We find that the prediction capability can indeed be improved, especially by removing old entries from the knowledge graph.
\end{enumerate}

The rest of the paper is organized as follows. We discuss related work on threat databases and knowledge graph approaches in Section~2. 
Next, in Section~3, we describe the creation of the threat knowledge graph and various methods for optimizing it. 
In Section~4, we detail the task of knowledge graph embedding, that is, translating the knowledge graph into vectors, and compare the performance of different embedding approaches.
In Section~5, we use the embedded knowledge graph for predicting hidden associations between CPE, CVE, and CWE, and evaluate the prediction capability with multiple metrics, both in closed-world and open-world settings.
In Section 6, we investigate methods to further optimize the knowledge graph for improving its prediction capability. We conclude the paper in Section~7.

\section{Related work}

There exists a rich literature on leveraging threat databases to derive insights about vulnerabilities and threats. 
In particular, the work in~\cite{li2017mining} applies data mining on the CVE and CWE databases to derive interesting characteristics  about software vulnerabilities, such as statistics of essential vulnerabilities. 
The work in~\cite{zhang2015predicting} uses historical data in threat databases for predicting the time until the next vulnerability will affect a specific software application.
Prediction models for a few specific vendors and applications are developed, and challenges to develop prediction models for more general cases are pointed out, such as inconsistency between CPE and CVE entries. Compared to~\cite{zhang2015predicting}, our work proposes a \emph{general} prediction model, based on knowledge graph embedding.

Leveraging threat databases is non-trivial since they are not always consistent and ready for use~\cite{zhang2015predicting}. 
Thus, efforts have been made to automate associations of entries in different threat databases.
The work in~\cite{waareus2020automated} proposes an automated way of extracting related CPE entries from a CVE description, to associate CPE and CVE entries. This is achieved by applying on CVE descriptions a natural language processing technique called named entity recognition (NER).
Automated association between CVE and CWE can be cast as a classification problem, since vulnerabilities can be viewed as instances of weaknesses in specific products or systems. 
The work in~\cite{aota2020automation} first vectorizes CVE descriptions, then classifies the vectorized descriptions into 19 CWE classes by using the random forest algorithm based on selected features.

A common limitation of the approaches mentioned above is that they treat each CVE entry separately. Therefore, implicit relations are difficult to identify.
Looking at the description of CVE-2021-21348 in Table~\ref{CVE-2021-21348}, one can easily see a relation between the keyword XStream and the {\tt cpe:a:xstream project:xstream:*:*} CPE entry. However, the description does not suggest any relation to CPE entries like {\tt cpe:a:oracle:banking platform:*:*}.
In comparison, our knowledge graph approach is able to extract not only explicit but also \emph{implicit} relations, by aggregating knowledge from all entries. 

Knowledge graphs have widely been used for reasoning over interrelated data for tasks such as recommendation~\cite{wang2019kgat} and question answering~\cite{huang2019knowledge}.
The work in~\cite{chen2020review} reviews approaches and applications for knowledge graph reasoning. As threat databases are interrelated through associations between their entries, knowledge graph approaches are suitable for understanding and investigating these associations. 
In order to complete various tasks such as link prediction, knowledge graphs are embedded onto a vector space.
Many embedding models have been proposed, including TransE~\cite{bordes2013translating}, DistMult~\cite{yang2014embedding}, and ComplEx~\cite{trouillon2016complex}. 
In~\cite{wang2017knowledge}, a comprehensive survey of the embedding approaches and the applications of embedded knowledge graphs is provided.
The quality of embedding is typically evaluated by rank-based metrics~\cite{bordes2013translating}.

The work in~\cite{wang2020analysis} produces a knowledge graph from CVE and CWE entries, and provides examples of querying the knowledge graph to analyze properties of vulnerabilities.
The work in~\cite{yuan2021predicting} produces a knowledge graph using multiple threat databases, including the CVE and the CWE, and uses the knowledge graph to predict hidden links, which correspond to associations within a threat database or between different databases.
Yet, the work in~\cite{yuan2021predicting} does not consider the CPE, though it is valuable for identifying threats based on products in the system. 
Moreover, evaluation of prediction is performed by splitting the knowledge graph into training and test sets under a closed-world assumption. Hence, it is unclear how the prediction model at a certain time performs for future data.
In our work, we also focus on finding relations between CVE entries to specific products in the CPE, to identify potential threats in products used in real systems.
Another unique contribution of our work is in the validation of our methodology using historical data to demonstrate its prediction capabilities (i.e., an open-world setting).

A shorter, preliminary version of this paper appeared in~\cite{shi2022uncovering}.
Compared to the work in~\cite{shi2022uncovering}, this manuscript includes the following significant new contributions:
\begin{itemize}
    \item We demonstrate the usefulness of our model for not only predicting associations between CVE and CPE entries, but also between CVE and CWE entries. The new corresponding results are reported in Sections~5 and~6.

    \item We conduct a comprehensive evaluation of different embedding models for the task at hand, including the TransE, DistMult, and ComplEx models.
    We evaluate them using rank-based metrics, and accordingly select the best model for predicting the associations.
    The comparison results of the embedding models are detailed in Table~\ref{table-comparison-aug2021} and Table~\ref{table-comparison-nov2022} in Section~4.
    We show that metrics obtained using TransE, with appropriate parameters, perform best, and consistently better than those provided in our preliminary work in~\cite{shi2022uncovering}.
    
    \item In~\cite{shi2022uncovering}, the model is evaluated by predicting new triples between August~4, 2021 and March~29, 2022 using data available before that time period.
    Here, we leverage recent updates of threat databases to investigate the prediction capability over time.
    Specifically, we find that the rank-based metrics \emph{improve} as more new associations are added to the test set, as shown in Table~\ref{table-prediction} and Fig.~\ref{pred_diff_test_set} in Section~5. This suggests that our prediction performance could be viewed as lower bounds on actual performance. 
    We also show that, as the knowledge graph gets updated, the prediction capability of the model stays strong, as shown in Table~\ref{table-prediction-consistency}.

     \item We evaluate the embedding and prediction results both for all triples in the graph as in~\cite{shi2022uncovering}, and only certain \emph{target triple types} (i.e., CVE\textrightarrow CPE and CVE\textrightarrow CWE triples). In Sections~4, 5, and~6, we evaluate rank-based metrics for these target triple types, and find that the performance of the model for these target triples is even better than the performance averaged over all triples.

    \item In Section~3.1,  we provide more details and examples  on how to generate the knowledge graph.

    \item We investigate and evaluate multiple approaches to further improve the prediction capability of the knowledge graph in Section~6. The approaches include removing old CPE and CVE entries, adding entries from the CAPEC database, and adding CVSS vectors.
    
\end{itemize}

\section{Threat knowledge graphs} 

In this section, we describe our methodology for producing a knowledge graph out of the CPE, CVE, and CWE databases. We first introduce the structure of the knowledge graph. 
We give examples to show how the original data from the databases are used to generate triples in the knowledge graph.
Alongside, we detail and justify optimizations performed on the knowledge graph.

\subsection{Structure of the knowledge graph} \label{kg_strut}

\begin{figure*}[t]
	\centering
	\includegraphics[width=0.9\linewidth]{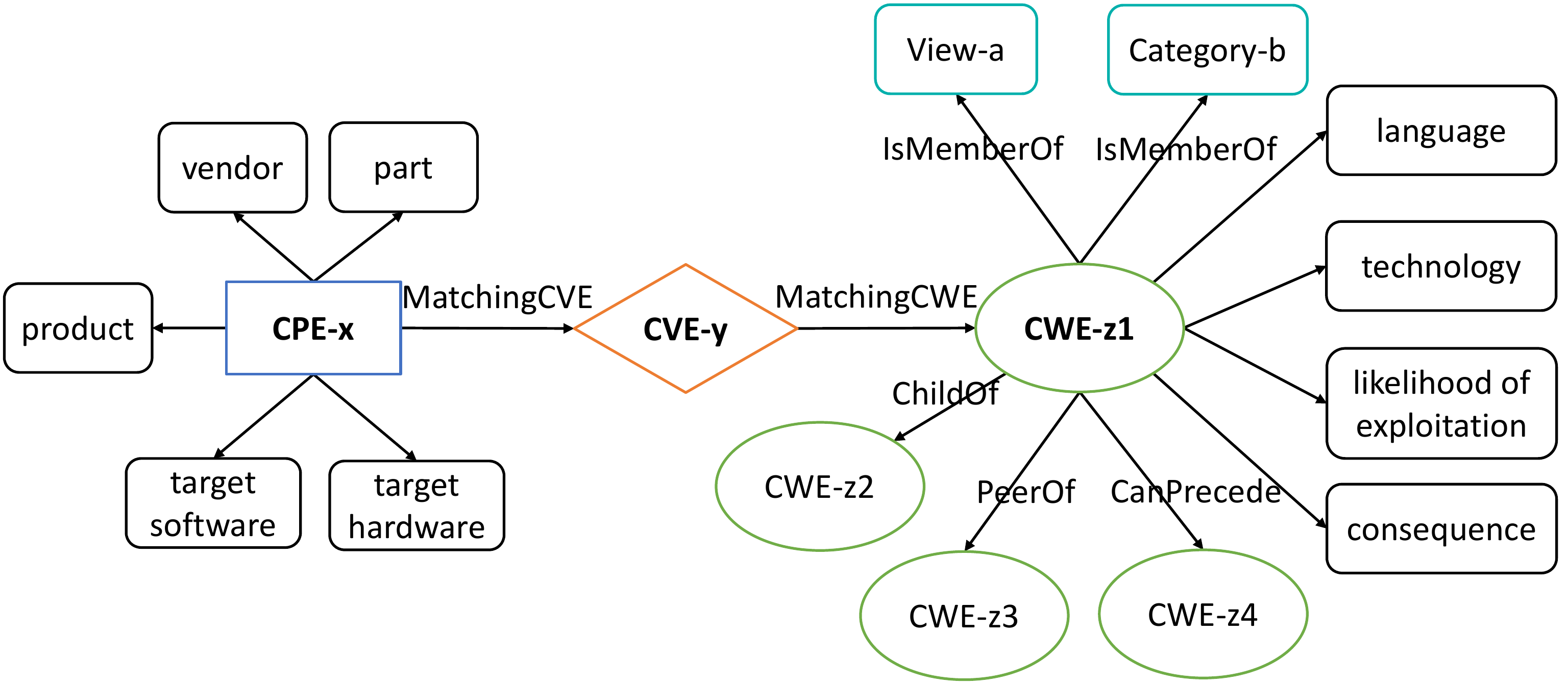}
	\caption{Complete structure of the threat knowledge graph. (a, b, x, y, z1-z4 represent the IDs of the corresponding entries.)}\label{kg}
\end{figure*}

A knowledge graph is represented with three sets $(E, R, T)$, namely the set of entities $E$, the set of relations $R$, and the set of triples $T$. Each triple $\tau \in T$ is of the form $\langle h,r,t \rangle$, where $h, r, t$ respectively correspond to the head entity,  the relation, and the tail entity. 
The structure of the knowledge graph is shown in Fig.~\ref{kg}.
Each threat database (CPE, CVE, and CWE) provides a list of \emph{entries}, which correspond to products, vulnerabilities, or weaknesses. In addition to the name, each entry has \emph{attributes} that provide further details, such as the vendor of a product. 
The entities in the knowledge graph are selected from the entries in the CPE, CVE, and CWE databases, as well as from the attributes of the entries, which are elaborated below. The arrows in Fig.~\ref{kg} point from the head entity to the tail entity.

There are three main types of entities, namely, product (CPE-x), vulnerability (CVE-y), and weakness (CWE-z), which are respectively based on the entries in the CPE, CVE, and CWE databases.
The views and categories in the CWE are also added as entities in the knowledge graph.
Each type of entries mentioned is added to the knowledge graph with a unique name (by directly using the name of the entry or making slight adjustments, if needed).

In addition to the names of the CPE and CWE entries,  key attributes are included in the knowledge graph as entities.
For each CPE entry, we include the following five attributes (if present): the part (application, hardware, or operating system), the vendor (e.g., Google), the product (e.g., Chrome), the target software (e.g., Node.js), and the target hardware (e.g., x86).
For each CWE entry, we include the following four attributes (if present): the language (e.g., JavaScript), the technology (e.g. web server), the likelihood of exploit (high/medium/low), and the consequence. The consequence describes which type(s) of security property are violated, for example, confidentiality and integrity.

The entities are connected by relations, as shown in Fig.~\ref{kg}. There exist three types of relations: (i)~the attribute of a node (e.g., the vendor of CPE-x); (ii)~existing association between CPE, CVE, and CWE entries as provided by the NVD (e.g., CPE-x is related to CVE-y, and CVE-y is related to CWE-z); (iii)~relations between weaknesses, views, and categories (e.g., CWE-z1 is a child of CWE-z2, and CWE-z1 is a member of View-a).

Next, we give a couple of examples that illustrate the process of constructing the knowledge graph.
When considering a CPE entry, we add to the knowledge graph an entity with a simplified name to represent the entry, then add corresponding entities for all applicable attributes (part, vendor, product, target software, and target hardware).
For example, consider the CPE entry ``cpe:2.3:a:vim:vim:5.6:*:*:*:*:*:*:*'', we first simplify its name as ``cpe:a:vim:vim:5.6:*:*'' by removing irrelevant information.
The details of the entry ``cpe:a:vim:vim:5.6:*:*'' is shown in Table~\ref{example_cpe}.
We have four corresponding entities: ``cpe:a:vim:vim:5.6:*:*'', ``a(part)'', ``vim(vendor)'', ``vim(product)''. We also have three triples reflecting the relations between the entities:
<cpe:a:vim:vim:5.6:*:*, HasPart, a(part)>, 
<cpe:a:vim:vim:5.6:*:*, HasVendor, vim(vendor)>, 
<cpe:a:vim:vim:5.6:*:*, HasProduct, vim(product)>.
The name ``cpe:a:vim:vim:5.6:*:*'' also suggests that the product version is 5.6.

\begin{table}[t]
    \centering
  \begin{tabular}
    {|p{0.25\textwidth}|p{0.08\textwidth}|p{0.08\textwidth}|p{0.08\textwidth}|p{0.17\textwidth}|p{0.18\textwidth}|}
    \hline
        Name & Part & Vendor & Product & Target Software & Target Hardware  \\ \hline
        cpe:a:vim:vim:5.6:*:* & a & vim & vim & * & *  \\ \hline
    \end{tabular}
    \caption{Example of a CPE entry.}
    \vspace{-0.5cm}
    \label{example_cpe}
\end{table}

When considering a CVE entry, we only add an entity with a corresponding name, without additional attributes.
Note that descriptions of CVE entries are unstructured text. Hence, these descriptions cannot be directly used as knowledge graph triples. As a result, we decided not to include them, and leave their incorporation into the threat knowledge graph as an area for future work.
Instead, we include associations from each CVE entry to CPE or CWE entries, which connect the three databases.
We use relations ``MatchingCVE'' for the mapping between a CPE and CVE entry, and ``MatchingCWE'' for the mapping between a CVE and CWE entry.
For instance, for the entry CVE-2021-29529 in Table~\ref{example_cve}, we add the triples
<cpe:a:google:tensorflow:2.4.1:*:*, MatchingCVE, CVE-2021-29529> and
<CVE-2021-29529, MatchingCWE, CWE-131>.

\begin{table}[t]
    \centering
    \begin{tabular}{|p{0.18\linewidth}|p{0.3\linewidth}|p{0.15\linewidth}|p{0.26\linewidth}|}
    \hline
        ID & Description & Matched CWE & Matched CPE  \\ \hline
        CVE-2021-29529 & TensorFlow is an end-to-end open source platform for ... & CWE-131; CWE-193 & cpe:a:google:tensorflow: 2.4.1:*:* \\ \hline
    \end{tabular}
    \caption{Example of a CVE entry.}
    \vspace{-0.5cm}
    \label{example_cve}
\end{table}

For a CWE entry that represents a weakness, we add its name and attributes (language, technology, likelihood of exploitation, consequence) as entities in the knowledge graph, similar to the way we treat CPE entries.
For example, for the CWE entry shown in Table~\ref{example_cwe}, we have 6 entities: ``CWE-1004'', ``Language-Independent'', ``Web Based'', ``Medium'', ``Confidentiality'', ``Integrity''. Note that each attribute may have multiple items, and each item will be a separate entity. The triples will be in the form <CWE-1004, HasTechnology, Web Based>, <CWE-1004, LikelihoodOfExploit, Medium>. A weakness may also be related to other weaknesses, with relations of the form ``ChildOf'', ``CanPreceed'', ``PeerOf'', and so on. Therefore, for the example CWE entry, we additionally include the triple <CWE-1004, ChildOf, CWE-732> in the knowledge graph.

\begin{table}[t]
    \begin{center}
    \begin{tabular}{|p{0.15\linewidth}|p{0.34\linewidth}|p{0.2\linewidth}|p{0.19\linewidth}|}
    \hline
        ID & Name & Related Weakness & Language  \\ \hline
        CWE-1004 & Sensitive cookie without \newline `HttpOnly' flag & ChildOf: 732 & Language-Independent  \\ \hline
    \end{tabular} \\
    \vspace{1em}
    \begin{tabular}{|p{0.15\linewidth}|p{0.14\linewidth}|p{0.3\linewidth}|p{0.29\linewidth}|}
    \hline
        ID  & Technology & Likelihood of Exploit & Consequence \\ \hline
        CWE-1004 & Web Based & Medium & Confidentiality; Integrity \\ \hline
    \end{tabular}
    \caption{Example of a CWE weakness entry.}
    \vspace{-0.5cm}
    \label{example_cwe}
    \end{center}
\end{table}

Besides weaknesses, CWE also includes views and categories based on the format shown in Table~\ref{example_view} and Table~\ref{example_cate}. The views and categories provide useful characteristics about weaknesses. Hence, these relations  are included in the knowledge graph. For the examples shown, this includes the triples <CWE-488, MemberOfCategory, Category-1217> and <CWE-1129, MemberOfView, View-1128>.

\begin{table}[t]
    \centering
    \begin{tabular}{|p{0.12\linewidth}|p{0.27\linewidth}|p{0.1\linewidth}|p{0.12\linewidth}|p{0.25\linewidth}|}
    \hline
        ID & Name & Type & Status & Has Member \\ \hline
        View-1128 & ``CISQ Quality Measures (2016)'' & Graph & Incomplete & CWE-1129; CWE-1130; CWE-1131; CWE-1132 \\ \hline
    \end{tabular}
    \caption{Example of a  CWE view entry.}
    \vspace{-0.5cm}
    \label{example_view}
\end{table}

\begin{table}[t]
    \centering
    \begin{tabular}{|p{0.16\linewidth}|p{0.26\linewidth}|p{0.1\linewidth}|p{0.36\linewidth}|}
    \hline
        ID & Name & Status & Has Member \\ \hline
        Category-1217 & ``User Session Errors'' & Draft & CWE-488; CWE-613; CWE-841 \\ \hline
    \end{tabular}
    \caption{Example of a CWE category entry.}
    \vspace{-0.5cm}
    \label{example_cate}
\end{table}


\subsection{Optimizing the knowledge graph}
\label{kg_opt}

The purpose of the threat knowledge graph is to extract  information from associations between the threat databases. 
Therefore, instead of using all the entries from the databases, we make the following optimizations: (i)~we merge CPE entries that have identical attributes, except for their version numbers; (ii)~we remove CPE and CVE entries that contain no association information (e.g., CPE entries that are not associated with any CVE entry).  
We discuss next why these two optimizations are needed to improve the quality of the knowledge graph.

\subsubsection{Merging CPE entries}

As of March~29, 2022, in total, the CPE has 858,409 entries, the CVE has 183,368 entries, while the CWE has 924 weakness entries, 326 category entries, and 46 view entries. The CWE only provides general information about threats, while the CPE and CVE are more specific. Hence, the size of the CWE is much smaller than those of the CVE and CPE. We also note that the CPE and CVE are regularly updated, while the CWE remains relatively stable.

In the CPE, the numbers of vendors and products (15,354 and 84,358, respectively) are significantly smaller than the total number of CPE entries, because many entries only differ by their version numbers. 
Among all the CPE entries, 742,868 are classified as ``application'', 77,151 are classified as ``operating system'', and 38,390 are classified as ``hardware''.
We find that the vast majority of the CPE entries are application-related, i.e., about 90\%. These application-related entries tend to have many more version numbers than the other two types. 
If we merge CPE entries that are identical except for their version numbers, the total number of CPE entries decreases from 858,409 to 89,895, which is only about 10\% of the total number of entries. In this manner, the size of the CPE component of the knowledge graph is significantly compressed.

We further find that, after merging, a small portion of application-related CPE entries tend to have many more version 
numbers than other entries. 
We note that over $85\%$ of the merged CPE entries have fewer than ten version numbers, but about $3\%$ of the entries have over 100 version numbers.
Without merging, entries with many version numbers would have a too large weight on the knowledge graph, making it difficult to extract information from other valuable entries. Thus, merging CPE entries results in a more balanced knowledge graph.

\subsubsection{Removal of unconnected CPE and CVE entries}

After the merging of CPE entries, the resulting knowledge graph has 89,895 CPE entities, 183,368 CVE entities, and 924 CWE entities. Yet, many entries are not connected to entries in other databases. 
Specifically, only 33,427 CPEs (37.2\%) are connected to CVEs, 149,582 CVEs (81.5\%) are connected to CPEs, 106,861 CVEs (58.3\%) are connected CWEs, and 288 CWEs (31.2\%) are connected to CVEs.

We define an ``unconnected'' entity as a CPE/CVE/CWE entry that is not associated to entries in other threat databases 
(e.g., a CVE entry that is not associated to either a CPE or a CWE entry).
The motivation behind creating a threat knowledge graph is to utilize relations between threat databases to discover new knowledge and get insights into common threats.
However, unconnected entities provide no information in terms of relations between threat databases.
Therefore, to increase the portion of valuable entries in the knowledge graph, we do not include unconnected CPE and CVE entries. 
Our evaluation in Section~\ref{kg_embedding} validates the effectiveness of this optimization.

\section{Threat knowledge graph embedding}
\label{kg_embedding}

\begin{figure*}[t]
    \centering \includegraphics[width=0.6\linewidth]{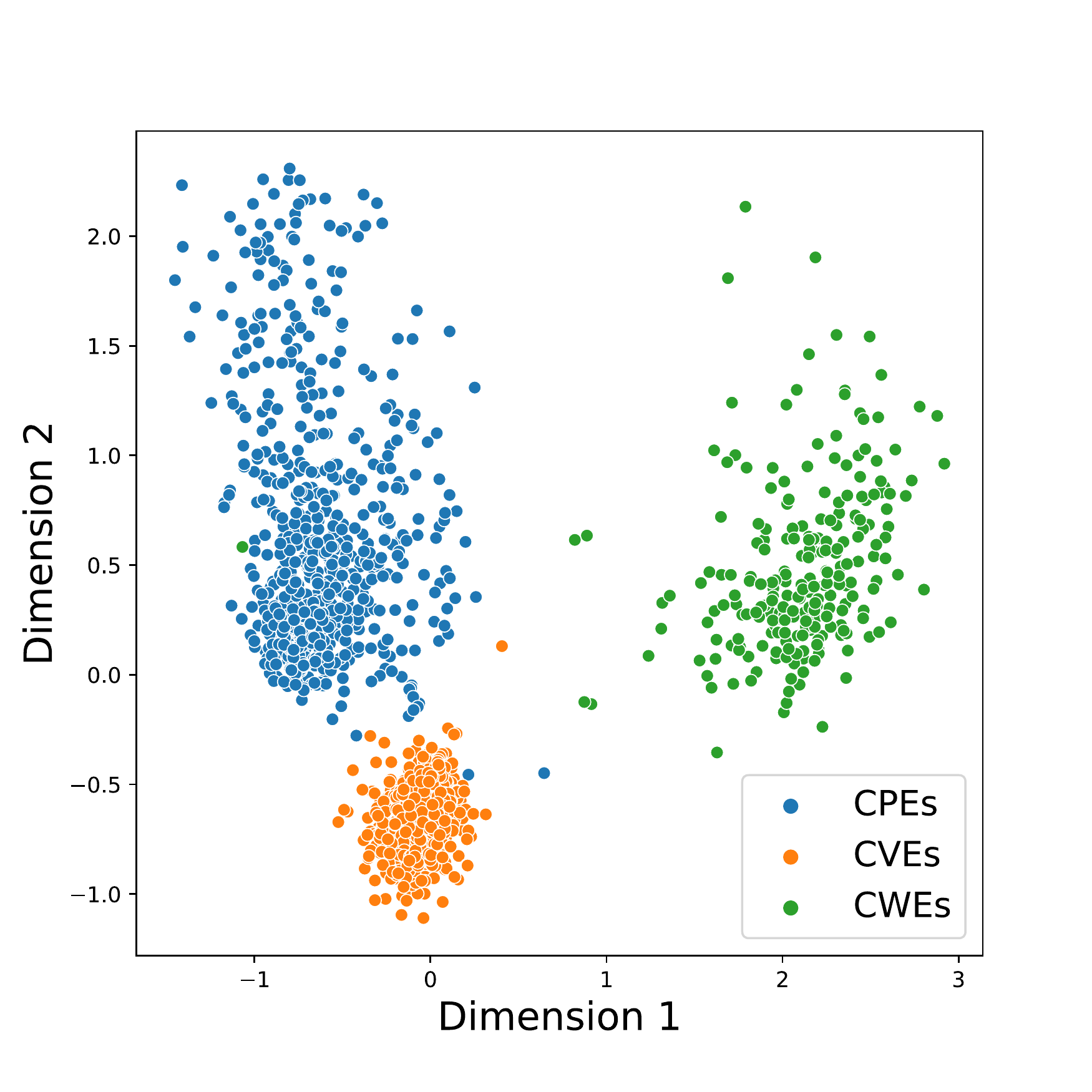}
    \caption{Illustration of knowledge graph embedding. The CPE, CVE, and CWE entries shown in Fig.~\ref{kg} are embedded as vectors in a 200-dimensional vector space, and then projected onto a 2-dimensional space by principal component analysis for illustration. 
    Note that the relations and attributes in Fig.~\ref{kg} are also embedded as vectors, but not shown here.
 }\label{illustration}
\end{figure*}

In this section, we undertake the task of knowledge graph embedding.
The goal of embedding is to translate entities and relations in the knowledge graph to vectors in a continuous vector space, as illustrated in Fig.~\ref{illustration}.
The embedded knowledge graph can then be used for various applications, including link prediction. 
We compare multiple state-of-the-art embedding models for our knowledge graph, and determine the model with the best performance (TransE in our case).

 

\subsection{Embedding model training}
\label{embedding_train}

In order to complete knowledge graph embedding, we need to adopt suitable embedding models. These models define how the entities and relations are translated into a vector space.
Multiple models are available, and they mainly differ by their \emph{scoring functions}~\cite{wang2017knowledge}. 
A scoring function $f(\langle h,r,t \rangle )$ assigns a score to each triple $\langle h,r,t \rangle$. A good scoring function generates high scores for positive triples (i.e., triples that exist in the real world) and low scores for negative triples.

TransE, DistMult, ComplEx, and HolE are among the state-of-the-art knowledge graph embedding models.
TransE~\cite{bordes2013translating} is based on translational distances in the vector space.
With TransE, the entities $h,t$ and the relation $r$ are all  translated into vectors $\mathbf{h}, \mathbf{t}, \mathbf{r}$ in real space $R^k$, such that $\mathbf{h}+\mathbf{r}$ can approximate $\mathbf{t}$ for the positive triples.
DistMult~\cite{yang2014embedding} represents a model that is based on semantic matching.
It uses the trilinear dot product to capture similarities between the triples. 
ComplEx~\cite{trouillon2016complex} extends DistMult in complex space with the Hermitian dot product, to better capture relations while maintaining a low computational complexity.
HolE~\cite{nickel2016holographic} is another semantic-matching-based model, but uses circular correlation.
The work in~\cite{hayashi2017equivalence} shows that ComplEx and HolE are equivalent for link prediction. Hence,  we focus on comparing the TransE, DistMult, and ComplEx  models in our evaluation.

In the training process, all the triples in the knowledge graph are considered as \emph{positive triples}, while \emph{negative triples} are the triples that do not exist in the knowledge graph.
Note that our knowledge graph only has positive triples, but negative samples are also needed for training, to help the embedding model differentiate between positive and negative triples. We follow the typical approach, where negative triples are generated from positive triples $\langle h,r,t \rangle$, by replacing $h$ or $t$ with another random entity in the knowledge graph. A generated negative triple is discarded if it belongs to  the knowledge graph (a step called \emph{filtering}).

The three embedding models used in our evaluations share a similar set of embedding parameters. For example, \emph{k} for the dimension of the embedding vector space, \emph{eta} for the number of negative triples generated against each positive one.
The model can also utilize the same loss functions, including pairwise loss, max-margin loss, and negative log-likelihood loss.
In order to make the results comparable, we test a large variety of combinations of the embedding parameters on each of the models, and compare the models with their most suitable parameters identified using grid search.

\subsection{Embedding model evaluation}
\label{embedding_eval}


We next evaluate the knowledge graph embedded by different models through standard metrics.
To evaluate the embedded knowledge graph, we select 10,000 triples from the entire knowledge graph (566,279 triples) as the test set (triples in this set are not used for training).
For each positive triple $\langle h,r,t \rangle$ in the test set, we perform the following: (i) generate all possible negative triples by replacing $h$ or $t$ with other entities; (ii) compute the scores of the positive triples and of the generated negative triples; (iii) \emph{rank} (order) the positive triples and the negative triples according to their scores (higher scores have lower rank). Positive triples in the test set should be ranked favorably compared to negative triples.

For quantitative evaluation of the embedding, the following metrics are commonly used (the mathematical definitions follow below)~\cite{bordes2013translating,yang2014embedding,trouillon2016complex}: (1)~the mean rank (MR), which computes the average of ranks of the positive triples; (2)~the mean reciprocal rank (MRR), which computes the reciprocal mean of ranks of the positive triples; and (3)~hits-at-N-score (Hits@N), which computes how many positive triples are ranked within the top~$N$ positions. Mathematically,
\begingroup
\allowdisplaybreaks
\begin{align*}
  MR&=\frac{1}{|T|}\sum_{i=1}^{T}rank_{\langle h,r,t \rangle_i}, \\
MRR&=\frac{1}{|T|}\sum_{i=1}^{T}\frac{1}{rank_{\langle h,r,t \rangle_i}}, \\
Hits@N&=\frac{1}{|T|} \sum_{i=1}^{T}\mathbf{1}\left[\ rank_{\langle h,r,t \rangle_i}\le N \right],
\end{align*}
\endgroup
where $T$ is a set of knowledge graph triples.
For a good embedding, we expect low MR, high MRR, and high Hits@N scores.

By computing the MR, MRR, and Hits@N scores for all knowledge graph triples in the test set, we get an overall evaluation of the embedded knowledge graph. 
The knowledge graph includes multiple types of relations (as illustrated by the arrows in Fig.~\ref{kg}).
We also focus on computing metrics for associations between CVE and CPE entries, as well as those between CVE and CWE entries.
In the test set, there are 5,612 CVE-CPE triples and 2,304 CVE-CWE triples, and we evaluate how well the embedding models learn these specific relations.
For CVE-CPE and CVE-CWE triples, we generate the negative triples by substituting the tail entity of the triples with the same type of candidates (\emph{filtering} out existing positive triples). 
For example, given a CVE-CWE triple, we generate negative triples by substituting the CWE entity with all the other CWE entities.
We denote the resulting set of positive and negative CVE-CWE triples as ``CVE\textrightarrow CWE triples''. The metrics for such triples can reflect how well we can predict the associated CWE entities for each CVE entity. Similarly, we have ``CVE\textrightarrow CPE triples'' to be evaluated.
Moreover, we evaluate the average metrics for all types of triples, where the negative triples are generated by substituting both sides of all triples in the test set with all the other candidate entities.
In this case, the resulting set of positive and negative triples are denoted as ``all triples''.

\begin{table}[t]
    \centering
    \begin{tabular}{|c|c|c|c|c|c|c|c|}
    \hline
    Triple type & Model & MRR & MR & Hits@20 & Hits@10 & Hits@3 & Hits@1 \\
    \hline
    \multirow{3}{*}{All} & TransE & \textbf{0.290} & \textbf{14206} & \textbf{0.430} & \textbf{0.391} & \textbf{0.314} & \textbf{0.238} \\
    & DistMult & 0.243 & 20091 & 0.376 & 0.332 & 0.259 & 0.196 \\
    & ComplEx & 0.255 & 21124 & 0.395 & 0.350 & 0.274 & 0.205 \\
    \hline
    \multirow{3}{*}{CVE\textrightarrow CPE} & TransE & \textbf{0.424} & \textbf{1643} & \textbf{0.626} & \textbf{0.573} & \textbf{0.468} & \textbf{0.345} \\
    & DistMult & 0.368 & 1924 & 0.539 & 0.490 & 0.397 & 0.302 \\
    & ComplEx & 0.392 & 1944 & 0.581 & 0.531 & 0.424 & 0.321 \\
    \hline
    \multirow{3}{*}{CVE\textrightarrow CWE} & TransE & \textbf{0.445} & \textbf{13} & \textbf{0.840} & \textbf{0.731} & \textbf{0.504} & \textbf{0.309} \\
    & DistMult & 0.315 & 228 & 0.518 & 0.444 & 0.338 & 0.244 \\
    & ComplEx & 0.291 & 218 & 0.533 & 0.450 & 0.324 & 0.207 \\
    \hline
    \end{tabular}
    \caption{Evaluation results of embedding using different embedding models for the threat knowledge graph. The knowledge graph uses data available on \textbf{August~4, 2021}. The metrics are computed for different types of triples, and TransE yields the best metrics in general. 
    ``All'' indicates computing the average metrics for all triples in the test set. 
    ``CVE\textrightarrow CPE'' (``CVE\textrightarrow CWE'') indicates that only CVE-CPE (CVE-CWE) triples are evaluated, and the negative triples are generated by replacing the CPE (CWE) entity.}
    \label{table-comparison-aug2021}
\end{table}

Table~\ref{table-comparison-aug2021} summarizes the evaluation results. 
For each embedding model, we have three groups of data, which correspond to the average metrics for all triples, the metrics for CVE\textrightarrow CPE triples, and the metrics for CVE\textrightarrow CWE triples in the test set.
The MR of 1643 implies that, on average, the TransE model ranks a positive CVE\textrightarrow CPE triple at the 1643-rd position against the negative triples generated from it (with about 30,000 negative triples). 
The Hits@3 score of 0.314 implies that the TransE model ranks positive triples within the top-3 positions 31.4\% of the time.

In general, we find that translational-distance-based models (TransE) perform better than semantic-matching models (DistMult, ComplEx) for our knowledge graph.
TransE performs the best among the three models in terms of all metrics.
The likely reason is that while the names of CPE entries in our knowledge graph consist of meaningful words (e.g., cpe:a:google:tensorflow:*:*) that provide additional information by themselves, the names of CVE and CWE entries are only ID numbers and do not imply properties about the entries.
If more semantic information about entries were added to the knowledge graph, ComplEx might have better metrics.
We also note that ComplEx is superior to DistMult for metrics that consider all triples and CVE-CPE triples only, but slightly worse for CVE-CWE triples only. The reason is the CPE side of the knowledge graph has rich semantics. As an improved version of semantic-matching models, ComplEx better utilizes the semantics, including learning relations between CPE and CVE entities. In contrast, the relations between CVE and CWE entities are less suitable for semantic-matching models, and ComplEx learns no better than DistMult.

We also note that all three models yield better metrics for the target CVE\textrightarrow CPE and CVE\textrightarrow CWE triples, compared to results averaged over all triples. 
Intuitively, this suggests that CVE\textrightarrow CPE and CVE\textrightarrow CWE triples are easier to learn than other types of triples in the knowledge graph.
Note that these two types of triples are exactly those we are interested in. Hence, our knowledge graph is well-suited for predicting the associations between CPE, CVE, and CWE entities.
We also observed that the knowledge graph predicts well relations in one direction (CVE\textrightarrow CPE and CVE\textrightarrow CWE), but metrics for triples in the opposite direction are generally worse.
The likely reason is that the number of CVE entries is significantly larger than the number of CPE and CWE entries. It is easier to make predictions starting from the side with more entries.

So far, we have computed the metrics based on a knowledge graph constructed with data available on Aug 4, 2021. In order to make sure that the comparison between the models truly reflects their performance difference, we perfom a similar evaluation based on a knowledge graph constructed with data available on Nov 1, 2022.
The results are shown in Table~\ref{table-comparison-nov2022}. We find that almost all metrics improve with the newer knowledge graph. Still, the relative performance between the three models is similar to that observed in Table~\ref{table-comparison-aug2021}. This suggests that our comparison results are consistent even though the knowledge graph grows over time.


\begin{table}[t]
\centering
\begin{tabular}{|c|c|c|c|c|c|c|c|}
\hline
Triple type & Model & MRR & MR & Hits@20 & Hits@10 & Hits@3 & Hits@1 \\
\hline
\multirow{3}{*}{All} & TransE & \textbf{0.303} & \textbf{15785} & \textbf{0.445} & \textbf{0.403} & \textbf{0.328} & \textbf{0.249} \\
& DistMult & 0.253 & 21442 & 0.394 & 0.350 & 0.271 & 0.204 \\
& ComplEx & 0.272 & 20633 & 0.415 & 0.372 & 0.294 & 0.220 \\
\hline
\multirow{3}{*}{CVE\textrightarrow CPE} & TransE & \textbf{0.465} & \textbf{2044} & \textbf{0.654} & \textbf{0.606} & \textbf{0.511} & \textbf{0.388} \\
& DistMult & 0.367 & 2158 & 0.556 & 0.508 & 0.400 & 0. 295 \\
& ComplEx & 0.414 & 2149 & 0.604 & 0.559 & 0.453 & 0.338 \\
\hline
\multirow{3}{*}{CVE\textrightarrow CWE} & TransE & \textbf{0.413} & \textbf{18} & \textbf{0.807} & \textbf{0.698} & \textbf{0.471} & \textbf{0.276} \\
& DistMult & 0.281 & 228 & 0.501 & 0.411 & 0.304 & 0.209 \\
& ComplEx & 0.282 & 163 & 0.545 & 0.453 & 0.308 & 0.195 \\
\hline
\end{tabular}
\caption{Evaluation results of embedding using different embedding models for the threat knowledge graph. The knowledge graph uses data available on \textbf{Nov 1, 2022}.
The TransE model still yields the best metrics in general, similar to the results in Table~\ref{table-comparison-aug2021}.
}
\label{table-comparison-nov2022}
\end{table}


Through comparison between the embedding models, we determine that TransE performs well for the task at hand. The metrics for the TransE model are sufficient for good prediction, as validated by our experiments in Section~\ref{kg_prediction}.  We therefore selected TransE as the embedding model for the prediction of unknown associations. Note that our knowledge graph has many entities with relatively sparse connections, making it hard to get metrics as good as those from benchmarking~\cite{akrami2020realistic}. 
Fine-tuning training parameters of the embedding model may help further improve these metrics.

Last, we investigate the influence of optimizing the knowledge graph under the TransE model, specifically removing unconnected CPE and CVE entities. The results are shown in Table~\ref{table-optimization}. After optimizing the knowledge graph, the MRR increases from 0.232 to 0.290. The Hist@N scores also increase, indicating that positive triples are more likely to be found in top-N results.
We conclude that removing unconnected CPE and CVE entries indeed improves the quality of our knowledge graph.

\begin{table}[t]
\centering
\begin{tabular}{|c|c|c|c|c|c|c|}
\hline
& MRR & MR & Hits@20 & Hits@10 & Hits@3 & Hits@1 \\
\hline
Unoptimized & 0.232 & 19642 & 0.363 & 0.326 & 0.261 & 0.178 \\
\hline
Optimized & \textbf{0.290} & \textbf{14206} & \textbf{0.430} & \textbf{0.391} & \textbf{0.314} & \textbf{0.238} \\
\hline
\end{tabular}
\caption{Evaluation results of embedding with TransE model, before and after removing unconnected CPE and CVE entities.}
\label{table-optimization}
\end{table}


\section{Prediction using threat knowledge graph}
\label{kg_prediction}

In this section, we use the threat knowledge graph to uncover hidden associations between threat databases, namely, associations between CVE and CPE, and those between CVE and CWE.
We evaluate the prediction results using rank-based metrics, as well as the precision, recall and F1-score, which we define in the next subsection.
Our results demonstrate that one can use historical data to correctly predict many associations of threats in the future.

\subsection{Prediction results and metrics}

We use entries from the CPE, CVE, and CWE databases available on August~4, 2021 to predict CVE-CPE and CVE-CWE associations that appeared from August~4, 2021 until March~29, 2022.
We note that our results should be viewed as lower bounds on performance, since some hidden vulnerabilities may be revealed after this period.

Toward this end, we produce a knowledge graph from the threat data available by August~4, 2021. Note that, using this approach, we can only predict associations between entries that already existed on August~4, 2021, and were not removed during the knowledge graph optimization.
Associations between the entries added after August~4, 2021 cannot be predicted, since the knowledge graph does not include their information.

From August~4, 2021 to March~29, 2022, there were 41,583 newly added CVE-CPE associations. 
We find that out of the newly added associations, 4,134 of them are between entities that already existed in the August~2021 version of the knowledge graph. 
In practice, we would like to retrieve those triples from a large set of candidates. To evaluate prediction in this scenario, we create a test set that includes both positive and negative triples, and the goal is to correctly predict them as positive or negative. We use the 4,134 new triples as positive triples in the test set.
On the other hand, there were 422 newly added CVE-CWE associations that are between entities that already existed, which serve as positive triples in the test set.
Following the approach detailed in Section~\ref{embedding_eval}, we generate the negative triples by substituting the CPE/CWE side of the positive triples with the same type of candidates.


The prediction is based on the embedding model of the knowledge graph. Each triple is assigned a score by the scoring function of the embedding model.
Scores are then normalized into estimated probabilities as the prediction output. 
A triple is predicted to be positive if its computed probability exceeds a \emph{prediction threshold} $\alpha$, where $0\le \alpha \le 1$.
A positive predicted triple means that the threat association (CVE-CPE or CVE-CWE) is currently hidden, but that we predict will be revealed in the future. 

In order to evaluate the prediction capability, we use two groups of metrics: (i) rank-based metrics, including MRR, MR, and Hits@N scores; (ii) precision, recall, and F1-score.
The rank-based metrics can indicate the prediction capability of the knowledge graph. 
If the positive triples in the test set are mostly ranked higher than the negative triples, then setting a proper threshold $\alpha$ can effectively separate positive triples from negative triples, and the top-ranked triples can be seen as the prediction results.

Furthermore, as this is essentially a prediction task, we use the \emph{precision}, \emph{recall}, and \emph{F1-score} to evaluate the prediction results. A high precision implies that a high fraction of the predicted positive triples is correct. A high recall implies that a high fraction of the positive triples is successfully identified. It is well-known that there exists a trade-off between the precision and recall metrics, resulting in a precision-recall curve. In order to combine and balance these two metrics, the F1-score, which is defined as the harmonic mean of the precision and the recall, is often used. A high F1-score indicates one is able to simultaneously achieve relatively high precision and high recall.

\subsubsection{Evaluation with rank-based metrics}

\begin{figure}[t]
	\centering
	\begin{subfigure}[b]{0.48\linewidth}
		\includegraphics[width=\linewidth]{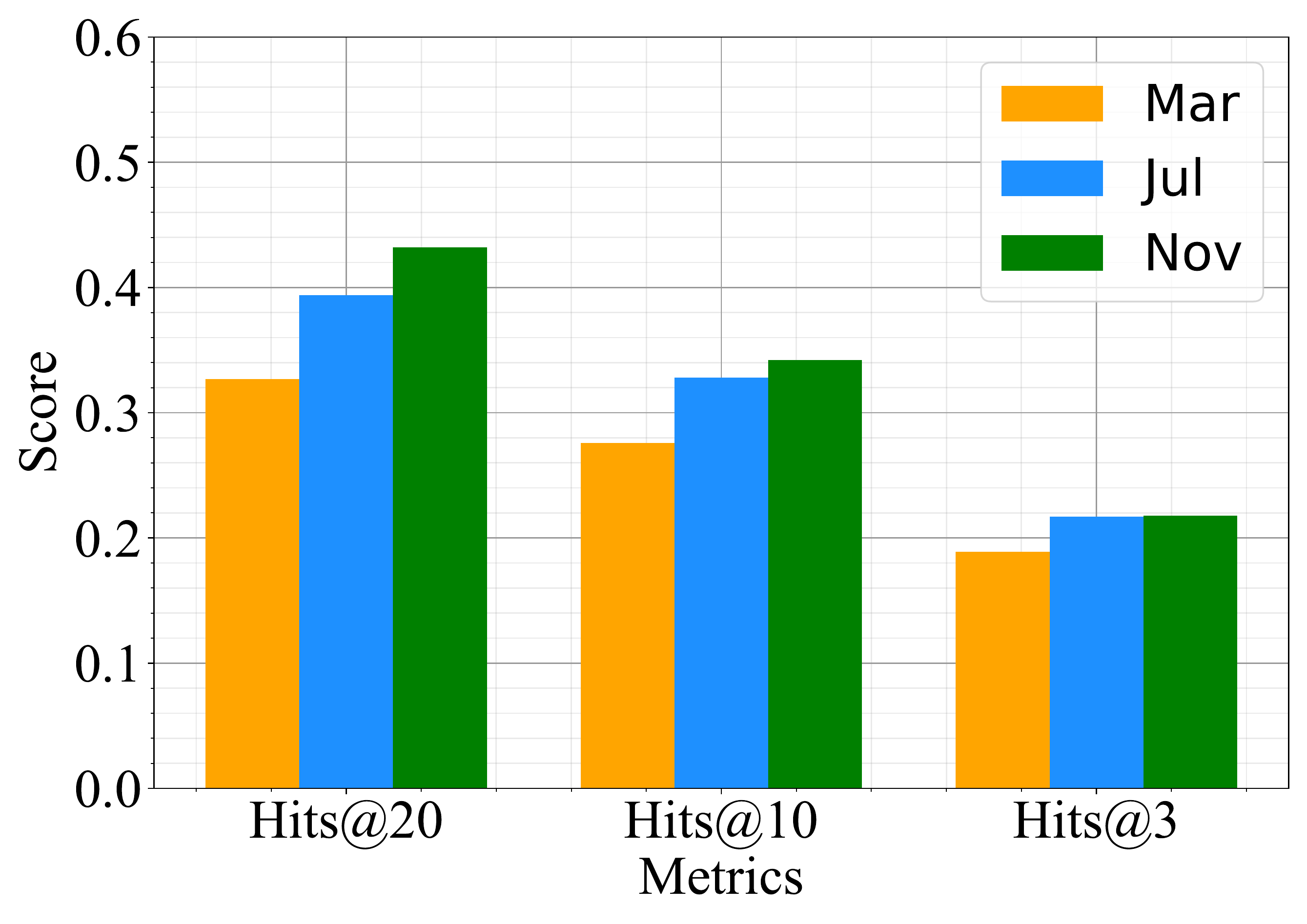}
		\caption{CVE\textrightarrow CPE}
	\end{subfigure}
	\begin{subfigure}[b]{0.48\linewidth}
		\includegraphics[width=\linewidth]{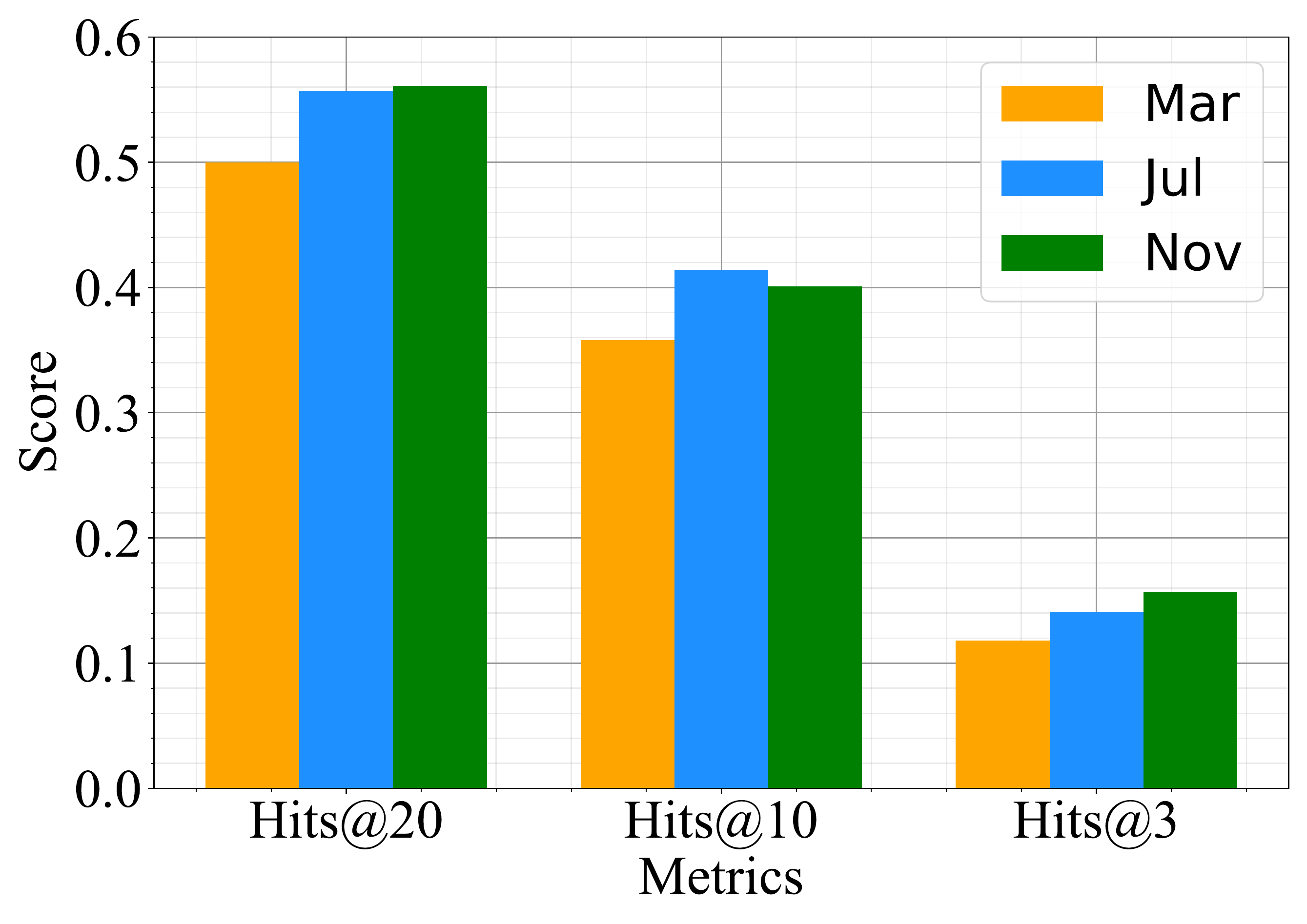}
		\caption{CVE\textrightarrow CWE}
	\end{subfigure}
	\caption{The Hits@N scores for predicting new associations, with the test set selected from different time periods.
    Note that only the end time of the periods is different (marked by different colors, and all in 2022), while the start time is Aug 2021 for all three groups.
    In general, the Hits@N scores increase as the test set includes more newly added triples.
    }
	\label{pred_diff_test_set}
\end{figure}

In Table~\ref{table-prediction}, we show the rank-based metrics when using the new triples to generate the test set. 
The results in Table~\ref{table-prediction} can be compared with the closed-world evaluation results in Table~\ref{table-comparison-aug2021}.
As expected, when predicting new triples in an open-world setting, the results are worse than when predicting triples in a closed-world setting (i.e., where the tested triples are from the knowledge graph itself). 
Nonetheless, the rank-based metrics suggest that the knowledge graph is able to provide useful predictions, which can significantly save manual work in finding potential triples.
Moreover, as explained earlier, metrics for new triples are  lower bounds, since some false positive triples turn out to be true positive in the future.
In order to validate this statement, we conduct evaluations with test sets covering different time periods (namely, new triples appearing until March~29, 2022, July~14, 2022, or November 1, 2022). 
We find that the metrics are improved in general when the test set covers a longer period, as shown in Table~\ref{table-prediction}.
For instance, when evaluating CVE\textrightarrow CWE prediction, the MRR increases from 0.143 to 0.168 if the test set covers new triples until November~2022 rather than March~2022.
The Hits@N scores also tend to increase, as illustrated in Fig.~\ref{pred_diff_test_set}.
Hence, this confirms that the actual prediction capability improves over time.


\begin{table}[t]
    \centering
    \begin{tabular}{|c|c|c|c|c|c|c|c|}
    \hline
           Triple type & Time range & MRR &  MR & Hits@20 & Hits@10 & Hits@3 & Hits@1 \\
         \hline
    \multirow{3}{*}{CVE\textrightarrow CPE} & Aug 2021 - Mar 2022 & 0.159 & 2068 & 0.327 & 0.276 & 0.189 & 0.091 \\
    & Aug 2021 - Jul 2022 & 0.186 & 1814 & 0.394 & 0.328 & 0.217 & 0.110 \\
    & Aug 2021 - Nov 2022 & 0.187 & 1277 & 0.432 & 0.342 & 0.218 & 0.104\\
    \hline
    \multirow{3}{*}{CVE\textrightarrow CWE} & Aug 2021 - Mar 2022 &  0.143 & 38 & 0.500 & 0.358 & 0.118 & 0.059 \\
    & Aug 2021 - Jul 2022 & 0.162 & 35 & 0.557 & 0.414 & 0.141 & 0.069 \\
    & Aug 2021 - Nov 2022 & 0.168 & 37 & 0.561 & 0.401 & 0.157 & 0.074 \\
    \hline
    \end{tabular}
    \caption{Evaluation results of predicting associations based on historical data, using the TransE model. 
    The time range specifies the time period during which the test set is selected. The metrics improve as the time range of the test set becomes wider, because the test set includes more newly added triples.}
    \label{table-prediction}
\end{table}

\subsubsection{Evaluation with precision, recall, and F1-score}

In this evaluation, for each unique CVE entry in new CVE-CPE triples, we generate 50 negative triples by substituting the CPE side.
As a result, we get 66,984 CVE-CPE triples in total for testing prediction, where 4,134 triples are positive, and 62,850 triples are negative.
Similarly, we generate 20 negative triples for each unique CVE entry in new CVE-CWE triples by substituting the CWE side.
We get 8,862 CVE-CWE triples in total for testing prediction, where 422 triples are positive, and 8,440 triples are negative.

\begin{figure}[t]
	\centering
	\begin{subfigure}[b]{0.48\linewidth}
		\includegraphics[width=\linewidth]{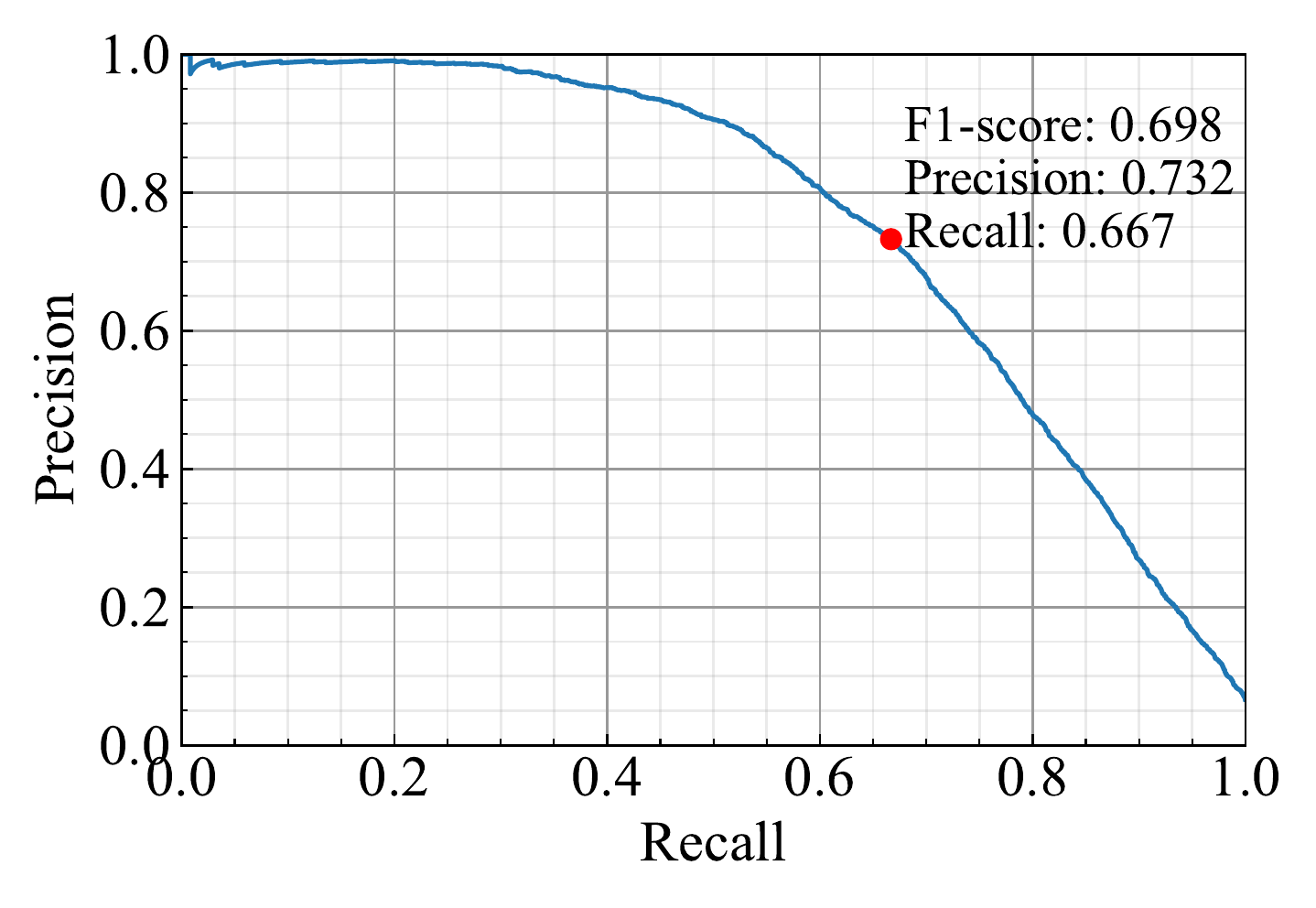}
		\caption{CVE\textrightarrow CPE}
	\end{subfigure}
	\begin{subfigure}[b]{0.48\linewidth}
		\includegraphics[width=\linewidth]{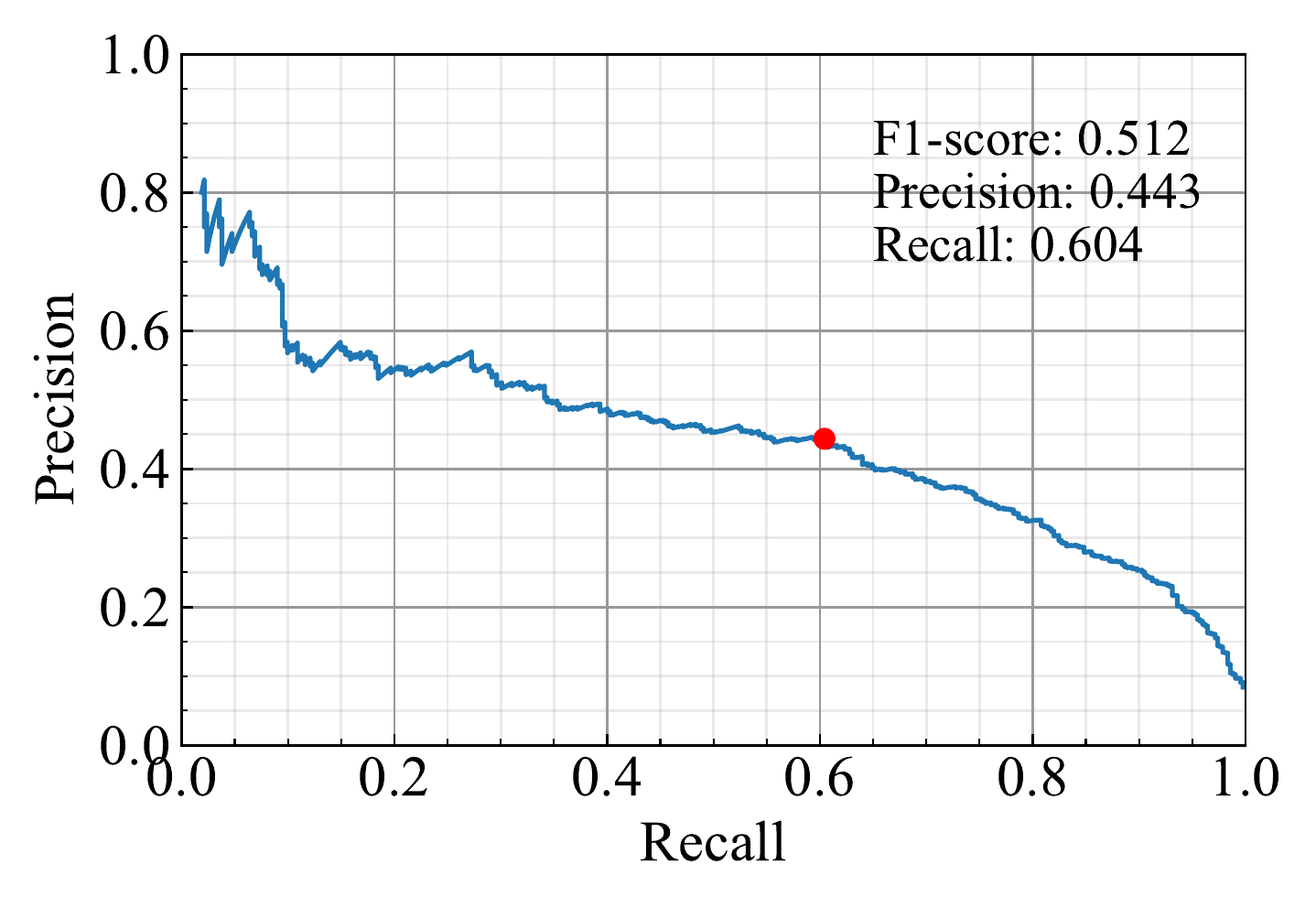}
		\caption{CVE\textrightarrow CWE}
	\end{subfigure}
	\caption{Precision-recall curve of predicting CVE-CPE associations using threat knowledge graph. The curve reflects the trade-off between precision and recall as the threshold $\alpha$ for positive prediction changes. When $\alpha$ decreases, the precision decreases and the recall increases. 
	A reference point based on the maximized F1-score is marked in red.}
	\label{pred_cpe2cve}
\end{figure}

We remind that our prediction model uses a prediction threshold $\alpha$.
The evaluation metrics change with $\alpha$ since only triples with estimated probability above $\alpha$ are predicted to be positive. For example, a high threshold $\alpha$ results in a higher precision but lower recall.
We depict the precision-recall curve in Fig.~\ref{pred_cpe2cve}, which reflects the trade-off between precision and recall as the threshold changes.
When $\alpha=0.981$, the F1-score for CVE-CPE prediction is maximized with a value of 0.681. The corresponding precision and recall are 0.775 and 0.606, respectively. This result means that about 77\% of the predicted associations are correct, while 60\% of unknown associations are retrieved.
Similarly, the F1-score for CVE-CWE prediction is maximized at 0.512, with the corresponding precision as 0.443 and recall as 0.604.
Note that the value of $\alpha$ that maximizes the F1-score only provides a reference. In practice, setting the proper value of $\alpha$ depends on the specific needs for precision and recall. For instance, a high value of $\alpha$ leading to a high precision might be preferred by a practitioner who wishes to have a reliable list of new associations that can easily be added to a threat database. On the other hand, a lower $\alpha$ leading to a high recall might be more suitable if a practitioner prefers to have an exhaustive candidate list of new associations which will be further verified manually.
Note that changing the number of negative triples may impact the precision (and thereby the F1-score), but not the recall. 

\subsection{Examples of prediction}

We next illustrate the prediction results through a few examples. 
The first example is CVE-2021-0144~\cite{nvd-cve-2021-0144} detailed in Table~\ref{CVE-2021-0144}. 
CVE-2021-0144 is a vulnerability related to Intel processors.
We explicitly separate its associated CPE entries into two groups: before August~4, 2021, and from August~4, 2021 through March~29, 2022. We predict CPE entries associated to CVE-2021-0144 in the second group, based on those in the first group. 
We only list a few known affected products (i.e., related CPE entries) due to the space limit. With $\alpha =0.96$, all the 465 affected products subsequently added to the NVD database are successfully predicted, with only 11 false positive predictions.

\begin{table}[t]
    \centering
    \begin{tabular}{|p{0.25\linewidth}|p{0.64\linewidth}|}
    \hline
        ID & CVE-2021-0144 \\ \hline
        Associated CWE & CWE-1188 \\ \hline
        Description & Insecure default variable initialization for the Intel BSSA DFT feature may allow a privileged user to potentially enable an escalation of privilege via local access. \\ \hline
        Associated CPE \newline by Aug 4, 2021 &  1) cpe:h:intel:core\_i5-7640x:*:*, \newline
        2) cpe:h:intel:core\_i9-10900x:*:*, \newline
        3) cpe:h:intel:xeon\_e5-1660\_v4:*:*, \newline
        4) cpe:h:intel:xeon\_platinum\_8274:*:*, \newline
        ...\\ \hline
        Associated CPE \newline after Aug 4,  2021  &  1) cpe:h:intel:atom\_c3000:*:*, \newline
        2) cpe:h:intel:core\_i3-6006u:*:*, \newline
        3) cpe:h:intel:core\_i9-11950h:*:*, \newline
        4) cpe:h:intel:xeon\_e-2124:*:*, \newline
        ... \\ \hline
    \end{tabular}
    \caption{Example of prediction results for CVE-2021-0144. 
    Using a threat knowledge graph generated with data available on August~4, 2021, we predict associations between CPE entries and the aforementioned vulnerability.
    Using a prediction threshold $\alpha = 0.96$, all the 465 affected products are successfully predicted with 11 false positive predictions.}
    \label{CVE-2021-0144}
\end{table}

A second, more complex example is CVE-2021-21348, detailed in Table~\ref{CVE-2021-21348} shown in the Introduction.
CVE-2021-21348 is a vulnerability related to a Java library called XStream. Apart from XStream itself, products that use XStream are also affected by this vulnerability, such as Debian Linux and Oracle SQL Server. In this example, the names of CPE entries associated to CVE-2021-21348 before August~4, 2021 are not directly related to the names of the CPE entries added afterward. Nonetheless, by aggregating relations between all CVE and CPE entries, the knowledge graph can discover the implicit relation between XStream and Oracle SQL Server. As a result, the association between {\tt cpe:a:oracle:mysql server:*:*} and CVE-2021-21348 can be uncovered accordingly. 
In general, by connecting various entries, the threat knowledge graph methodology is able to uncover many  associations based on implicit relations, in contrast to prediction approaches that consider each entry separately.

\subsection{Usefulness of prediction capability over different periods}

The prediction results above are based on the knowledge graph using data available on August~4, 2021. 
As the knowledge graph changes over time, in this subsection, we investigate whether the prediction capability remains useful over different time periods, using rank-based metrics.
In order to do so, we generate another threat knowledge graph, using data on March~29, 2022, and evaluate how well we can predict associations with it.

We generate two knowledge graphs, the first one using data available on August~4, 2021, and the second using data available on March~29, 2022.
After the optimization introduced in Section~\ref{kg_opt}, 
the knowledge graph using data on August~4, 2021 has 28,033 CPE entities, 237 CWE entities, 108,128 CVE entities connected to CPEs, and 84,307 CVE entities connected to CWEs.
In comparison,
the knowledge graph using data on March~29, 2022 has 33,427 CPE entities, 288 CWE entities, 149,582 CVE entities connected to CPEs, and 106,861 CVE entities connected to CWEs.

For each knowledge graph, we generate a test set from the new triples added in a few following months.
We also keep the time span of the two test sets roughly the same (due to limitations of the data we have, we cannot make them exactly the same).
Specifically, for the knowledge graph using data by August~4, 2021, the test set is generated from the new triples from August~4, 2021 to March~29, 2022;
For the knowledge graph using data by March~29, 2022, the test set is generated from the new triples from March~29, 2021 to November 1, 2022.
During March~29, 2021 to November 1, 2022, 7,597 new CVE-CPE associations and 1,041 new CVE-CWE associations are added that are between entities that already existed in the knowledge graph.
We use the MRR, MR, and Hits@N scores to evaluate how well the triples in the test sets can be predicted.

The evaluation results of the two knowledge graphs are shown in Table~\ref{table-prediction-consistency}.
We find that the knowledge graph using data on March~2022 predicts new triples in the following months better than the knowledge graph using data on August~2021.
Since the two test sets are generated from different time periods, we can expect the metrics to vary accordingly. 
Nonetheless, the MRR, Hits@10, Hits@3, and Hits@1 scores are roughly on the same level.
Thus, we can conclude that a knowledge graph generated by our approach can consistently predict threat associations uncovered in the future.
Moreover, we can expect the prediction metrics similar to those in Table~\ref{table-prediction-consistency}, if not better.

\begin{table}[t]
    \centering
        \begin{tabular}{|c|c|c|c|c|c|c|c|}
    \hline
           Triple type & Knowledge graph date & MRR &  MR & Hits@20 & Hits@10 & Hits@3 & Hits@1 \\
         \hline
    \multirow{2}{*}{CVE\textrightarrow CPE} & Aug 4, 2021 & 0.159 & 2068 & 0.327 & 0.276 & 0.189 & 0.091 \\
    & Mar 29, 2022 & 0.185 & 748 & 0.484 & 0.369 & 0.208 & 0.094 \\
    \hline
    \multirow{2}{*}{CVE\textrightarrow CWE} & Aug 4, 2021 &  0.143 & 38 & 0.500 & 0.358 & 0.118 & 0.059 \\
    & Mar 29, 2022 & 0.175 & 38 & 0.552 & 0.383 & 0.185 & 0.073 \\
    \hline
    \end{tabular}
    \caption{Comparison of the prediction metrics of two knowledge graphs.
    For the knowledge graph using data by Aug 4, 2021, the test set is generated from the new triples from Aug 4, 2021 to Mar 29, 2022.
    For the knowledge graph using data by Mar 30, 2022, the test set is generated from the new triples from Mar 30, 2021 to Nov 1, 2022.
    The metrics suggest that our approach can consistently make useful predictions.
    }
    \label{table-prediction-consistency}
\end{table}

\section{Improving the threat knowledge graph}

In this section, we further optimize the threat knowledge graph for predicting CVE-CPE and CVE-CWE associations. The approaches include and removing old entries and adding more useful threat information (such as the CAPEC database and CVSS vectors). 
We first provide details on each of these three approaches, and then evaluate their effectiveness using rank-based metrics.

\begin{figure*}[t]
	\centering
	\includegraphics[width=0.9\linewidth]{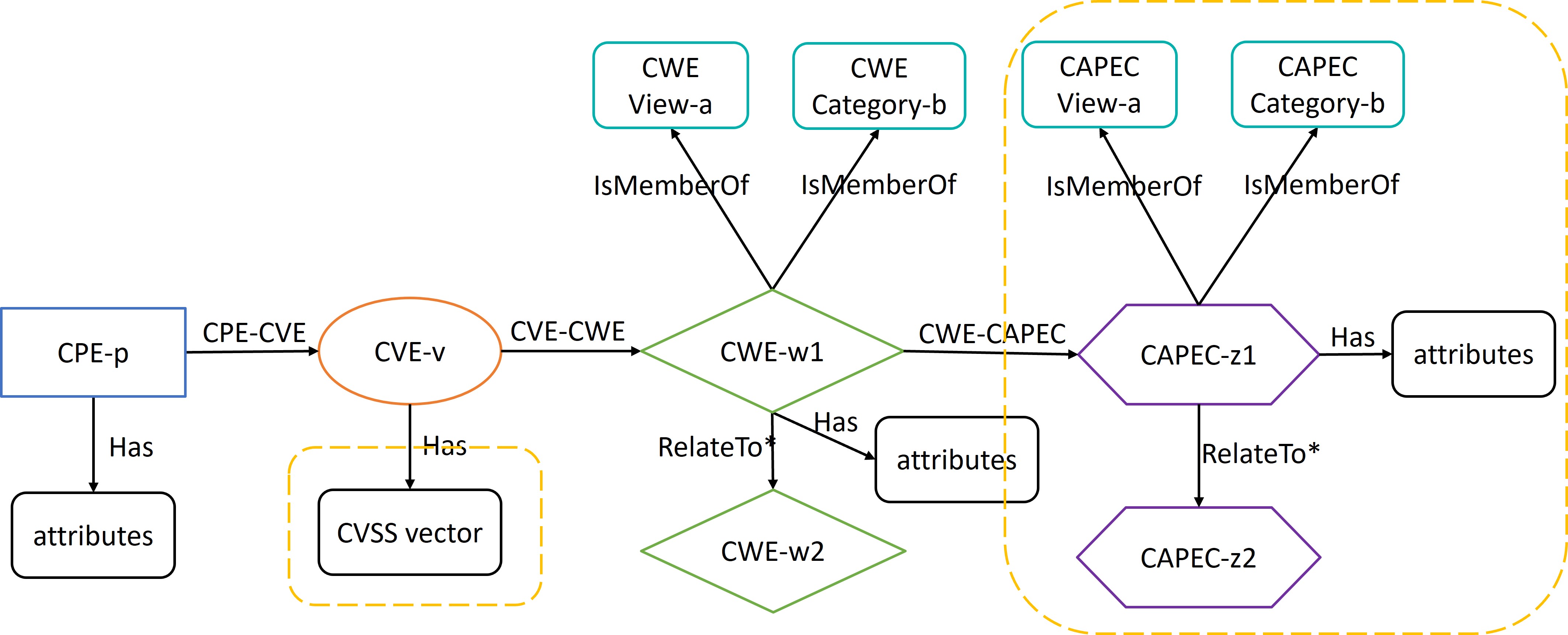}
	\caption{Structure of the threat knowledge graph with CAPEC and CVSS vectors. The parts within the yellow borders represent the new data added to the knowledge graph.}\label{kg_extend}
\end{figure*}

\subsection{Removal of old entries}

By observing the dates of creating CVE entries and associating them with CPE entries, we find that most associations are added within two years after a CVE entry is created.
Furthermore, over $94\%$ of the associations added between August~4, 2021 and July~14, 2022 only involve  CVE entries created after 2016.
As a result, old entries tend to become obsolete if we only care about predicting future associations.
Thus, we propose to improve the prediction capability of the knowledge graph, by removing old entries before a certain date.

Specifically, we first remove all the CVE entries that were created before 2016. The removal of old CVE entries results in new unconnected CPE entries, and we remove them as well. Since the CWE entries are relatively stable and interconnected with each other, we do not modify the CWE side of the knowledge graph. 
As a result, the new threat knowledge graph has much fewer CPE and CVE entries. The total number of triples decreases by $34.8\%$, from 380,138 to 247,705.
We expect that the remaining entities and relations are more closely related to the associations added in the future. 

\subsection{Adding the CAPEC database}

The Common Attack Pattern Enumeration and Classification (CAPEC) database~\cite{capec} enumerates known attack patterns employed to exploit weaknesses from the adversary's perspective.
The CAPEC entries are associated with the CWE entries, such that we can add them to provide more useful information.
We add the CAPEC entries as well as their attributes to the knowledge graph as entities. These new entities introduce new relations, including relations between CAPEC entries and their attributes, and associations between CAPEC entries and CWE entries. As a result, we have richer information on the CWE side of the knowledge graph, which could be useful for predicting the associations between CPE, CVE, and CWE entries.
The CAPEC part of the knowledge graph is illustrated within the yellow border on the right in Fig.~\ref{kg_extend}.

\subsection{Adding CVSS vectors}

In the threat knowledge graph, no attributes of CVE entries are included. 
We can enrich the information about the CVE entries by using the CVSS vectors provided by NVD.
Each CVE entry is associated with a CVSS vector, which evaluates the vulnerability in three metric groups. 
Each CVSS metric takes a predefined value. For example, the attack vector (AV) metric is of the form network (N), adjacent (A), local (L), or physical (P), and the attack complexity (AC) metric is of the form low (L) or high (H).
Thus, by including CVSS vectors, we add more information on CVE entries to the knowledge graph.
This information should help the embedding models better learn the similarities between different CVE entries.
The CVSS part of the knowledge graph is illustrated within the yellow border on the left in Fig.~\ref{kg_extend}.

\subsection{Evaluation of the approaches}

Next, we evaluate the effectiveness of the three approaches introduced above.
For each approach, we generate a new threat knowledge graph, and we compare them with the ``base'' knowledge graph detailed in Section~3.
We conduct both closed-world evaluations, where the test set is split from the knowledge graph itself, and open-world evaluations, where the test set is from the new triples added afterward.

The evaluation results are shown in Table~\ref{table-improve-closed} and Table~\ref{table-improve-open}.
For closed-world evaluations, we find that the knowledge graph using data after 2016 and the knowledge graph with CAPEC yield similar metrics to those of the base knowledge graph.
On the other hand, the knowledge graph with CVSS vectors has better embedding metrics on average for all types of triples, but still does not outperform the base knowledge graph when it comes to CVE\textrightarrow CPE and CVE\textrightarrow CWE triples.
Overall, the closed-world evaluation results suggest that the proposed approaches will not let the knowledge graph learn existing CVE-CPE and CVE-CWE associations better.

Although the closed-world embedding metrics are not significantly improved by the proposed approaches, the open-world evaluations show that, some of the approaches indeed help predict the associations that will be revealed and added in the future.
For CVE\textrightarrow CPE triples, removing old entries before 2016 brings the most significant improvement. For instance, the MRR increases by $35\%$, and the Hits@10 score increases by $34\%$.
For CVE\textrightarrow CWE triples, both removing the old entries and adding CVSS vectors significantly improve the embedding metrics.
Adding the CAPEC database, on the contrary, does not improve the prediction performance, possibly due to the fact that CWE-CAPEC associations are not strongly related to CVE-CPE and CVE-CWE associations. Nonetheless, adding CAPEC provides us with the bonus of predicting new CWE-CAPEC associations.
The likely reason that adding CVSS vectors works better for CVE\textrightarrow CWE triples is that associations between CVE and CWE entries are much denser than those between CPE and CVE entries (e.g., a CWE entry can be associated with hundreds of CVE entries).

\begin{table}[t]
    \centering
    \begin{tabular}{|c|c|c|c|c|c|c|c|}
    \hline
    Triple type & Knowledge graph & MRR &  MR & Hits@20 & Hits@10 & Hits@3 & Hits@1 \\
    \hline
    \multirow{2}{*}{All} & Base & 0.290 & 14206 & 0.430 & 0.391 & 0.314 & 0.238 \\
    & After 2016  & 0.297 & 7340 & 0.446 & 0.400 & 0.320 & 0.242 \\
    & With CAPEC & 0.277 & 13507 & 0.421 & 0.378 & 0.300 & 0.224 \\
    & With CVSS & \textbf{0.391} & \textbf{4334} & \textbf{0.493} & \textbf{0.469} & \textbf{0.430} & \textbf{0.339} \\
    \hline
    \multirow{4}{*}{CVE\textrightarrow CPE} & Base & 0.424 & 1643 & \textbf{0.626} & \textbf{0.573} & \textbf{0.468} & 0.345 \\
    & After 2016 & \textbf{0.426} & \textbf{1278} & 0.620 & 0.566 & 0.460 & \textbf{0.354} \\
    & With CAPEC & 0.420 & 1810 & 0.610 & 0.563 & 0.460 & 0.346 \\
    & With CVSS & 0.380 & 1565 & 0.602 & 0.549 & 0.438 & 0.290 \\
    \hline
    \multirow{4}{*}{CVE\textrightarrow CWE}  & Base & \textbf{0.445} & 13 & 0.840 & \textbf{0.731} & \textbf{0.504} & \textbf{0.309} \\
    & After 2016 & 0.415 & 17 & 0.804 & 0.686 & 0.464 & 0.285 \\
    & With CAPEC & 0.411 & 15 & 0.813 & 0.701 & 0.451 & 0.280 \\
    & With CVSS & 0.441 & \textbf{11} & \textbf{0.860} & 0.719 & 0.485 & 0.307 \\
    \hline
    \end{tabular}
    \caption{Closed-world evaluation results of embedding using TransE model. Four knowledge graphs are compared: the base knowledge graph introduced in Section~3, the graph using data after 2016, the graph with CAPEC, and the graph with CVSS vectors.
    The best metrics for each triple type are marked by bold text.
    }
    \label{table-improve-closed}
\end{table}

\begin{table}[t]
    \centering
    \begin{tabular}{|c|c|c|c|c|c|c|c|}
    \hline
    Triple type & Knowledge graph & MRR &  MR & Hits@20 & Hits@10 & Hits@3 & Hits@1 \\
    \hline
    \multirow{4}{*}{CVE\textrightarrow CPE} & Base & 0.159 & 2068 & 0.327 & 0.276 & 0.189 & 0.091 \\
    & After 2016 & \textbf{0.215} & \textbf{1488} & \textbf{0.429} & \textbf{0.341} & \textbf{0.241} & \textbf{0.146} \\
    & With CAPEC & 0.168 & 2214 & 0.343 & 0.300 & 0.205 & 0.095 \\
    & With CVSS & 0.170 & 1712 & 0.324 & 0.272 & 0.202 & 0.106 \\
    \hline
    \multirow{4}{*}{CVE\textrightarrow CWE}  & Base & 0.143 & 38 & 0.500 & 0.358 & 0.118 & 0.059 \\
    & After 2016 &  \textbf{0.203} & 37 & 0.597 & 0.432 & \textbf{0.219} & \textbf{0.098} \\
    & With CAPEC & 0.147 & 39 & 0.517 & 0.351 & 0.140 & 0.052 \\
    & With CVSS & 0.192 & \textbf{36} & \textbf{0.628} & \textbf{0.443} & 0.194 & 0.083 \\
    \hline
    \end{tabular}
    \caption{Open-world evaluation results of predicting associations based on historical data, using TransE model. Four knowledge graphs are compared: the base knowledge graph introduced in Section~3, the graph using data after 2016, the graph with CAPEC, and the graph with CVSS vectors.
    The best metrics for each triple type are marked by bold text.
    }
    \label{table-improve-open}
\end{table}

\section{Conclusion}

In this work, we introduced a methodology for predicting future associations between products, vulnerabilities, and weaknesses, using threat knowledge graphs. 
We explained how to generate and optimize such threat knowledge graphs, and how to embed them into a vector space. 
Through comparison between multiple embedding models, we found that the TransE model performs the best for our task.
We validated our approach with both closed-world and open-world evaluations. 
In open-world evaluation, we demonstrated good rank-based metrics on the associations revealed between August~4, 2021 and Nov~1, 2022, based on historical data by August~4, 2021.
This suggested that we can make good predictions on future associations with the help of the knowledge graph.
Moreover, we showed that the prediction capability of our knowledge graph improves over time
and is further improved with judicious optimizations, such as removing outdated entries.
One can utilize the predicted associations between entries of different threat databases in various ways. For example, if a specific product is used in a system, a threat modeler can list both existing and predicted vulnerabilities associated with the product. Our prediction methodology can also help prioritize manual verification of potential vulnerabilities of products.

Directions for future work include developing an embedding model for threat knowledge graphs that is specifically tailored to threat databases and could further improve prediction performance. 
Another direction is to investigate more approaches to optimize the knowledge graph, for example, incorporating the descriptions of CVE entries into the knowledge graph.
The CVE descriptions indeed provide rich information on the vulnerabilities, but the challenge is that they are unstructured, and require different techniques (e.g., NLP) to convert them into knowledge graph triples.

\section*{Acknowledgements}
This work was supported in part by Honda
Research Institute Europe GmbH and BU Hariri Institute Research Incubation Award (2020-06-006), the Boston University Red Hat Collaboratory (award \#
2022-01-RH03), and by the US National Science Foundation under grants CNS-1717858, CNS-1908087, CCF-2006628, and EECS-2128517.

\bibliographystyle{ACM-Reference-Format}
\bibliography{references}

\end{document}